\documentclass[twocolumn,twoside,prd,floatfix,letterpaper]{revtex4}

\usepackage{ifpdf}
\ifpdf 
  \usepackage[pdftex]{graphicx} 
\else 
  \usepackage{graphicx} 
\fi
\usepackage{color}
\usepackage{enumitem}
\usepackage{makecell}
\usepackage[colorlinks,hyperindex]{hyperref}

\def \lastDataDate {June 30, 2021}  
\newcommand{\lastDataDateSpecific}[1]{#1}
\def \suspiciousReturnFigure {Figure~\ref{fig:SuspiciousReturns}}
\def \citeKnuteson {\cite{knuteson2016information,knuteson2018wealth,knuteson2019celebrating,knuteson2020strikingly}}
\def \citeAllOvernightIntradayLiterature {\cite{knuteson2016information,knuteson2018wealth,knuteson2019celebrating,knuteson2020strikingly,cooper2008return,lachance2015night,qiao2020overnight,kelly2011returns,berkman2012paying,branch2012overnight,lou2019tug,bogousslavsky2021cross,lachance2020etfs}}
\def \TheNYT {\textit{The New York Times}}
\def \TheWSJ {\textit{The Wall Street Journal}}
\def \TheWashPost {\textit{The Washington Post}}
\def \FT {\textit{Financial Times}}

\begin{document} 

\title{They Chose to Not Tell You}
\author{Bruce Knuteson}
\noaffiliation

\begin{abstract}
The world's stock markets display a strikingly suspicious, decades long pattern of overnight and intraday returns that nobody (other than us) has plausibly explained and that nobody (other than us) has clearly and persistently alerted you to.  We use correspondence on this topic over the past five years to show that the silence of others on this issue does not arise from their having a good reason to believe this pattern is fine.  Separately, and regardless of whether this pattern turns out to be fine, we have documented that people in a position to alert you to the presence of strikingly suspicious return patterns in the world's stock markets that nobody can innocuously explain are aware of this issue, have no good reason to believe it is not a problem, and chose to not tell you.
\end{abstract}

\maketitle

\section{Goals\label{sec:Goals}}

This is the fifth -- and, for the fifth time, hopefully final -- article in what has become a roughly annual series~\citeKnuteson\ pointing out the existence of strikingly suspicious overnight and intraday return patterns in the world's stock markets, falsifying popular attempted explanations, and repeating the stunningly obvious point that strikingly suspicious return patterns in financial markets with no apparent innocuous explanation should be viewed as a serious problem.

To get in the mood, turn the page and get an eyeful of \suspiciousReturnFigure.

This article has two goals.

\subsection{To alert you to a strikingly suspicious return pattern and show why the silence of others does not mean everything is fine\label{sec:Goals:Alert}}

We have previously provided the only plausible explanation so far advanced for the strikingly suspicious overnight and intraday return patterns in \suspiciousReturnFigure~\cite{knuteson2019celebrating,knuteson2018wealth,knuteson2016information} and have clearly explained why other popular attempted explanations are not plausible~\cite{knuteson2020strikingly}.  Noting that we appear to be the only people publicly concerned about \suspiciousReturnFigure, we have also explicitly addressed the silence of others~\cite{knuteson2020strikingly} by explaining relevant barriers to knowledge on this issue and the strong incentives knowledgeable people have to be silent~\cite{knuteson2019celebrating,knuteson2018wealth}.

The logic leading a person observing the silence of others to conclude everything is fine contains an important intermediate step worth stating explicitly:  If others are silent about a seemingly obvious problem, then they must have a good reason for being unconcerned; and if these others have a good reason for being unconcerned, then everything is fine.  The only valid way to get from ``the silence of others'' to ``everything is fine'' is through this intermediate step.  A bunch of people with no good reason to think everything is fine is not a good reason to think everything is fine.

After briefly reviewing \suspiciousReturnFigure\ in Section~\ref{sec:StrikinglySuspicious}, we directly attack this intermediate step in Section~\ref{sec:Chose}.  Using six email threads that are representative of hundreds on this topic over the past five years, we show that the silence of others does not mean they have a good reason for being unconcerned.  The silence of others is therefore not evidence that \suspiciousReturnFigure\ is fine.  

\subsection{To document an extraordinary failure of information transfer \label{sec:Goals:Document}}

Errors in information transfer come in two flavors:  you can receive information that turns out to be wrong, and you can fail to receive important information that turns out to be right.  Notable cases of the latter range from Enron~\cite{swartz2003power} and Madoff~\cite{markopolos2010no}, where investors should have been (but were not) alerted to ongoing financial fraud, to cigarettes~\cite{proctor2012golden} and certain prescription painkillers~\cite{keefe2021empire}, where consumers and patients should have been (but were not) adequately alerted to the addictiveness and potential harmfulness of profitable products.  In these cases and others, many people have unnecessarily suffered because the institutions and personal incentives they counted on to get the information they needed failed them.

The second purpose of this article is to document an extraordinary failure in the institutions and personal incentives we currently rely on for the transfer of important information.  Setting completely to the side the issue of whether \suspiciousReturnFigure\ turns out to be the problem we strongly believe it to be, we clearly show five facts:  (i) the world's stock markets display a stunning pattern of overnight and intraday returns, (ii) many of the people you reasonably rely on to bring such an issue to your attention -- including financial regulators, journalists, and academic economists -- are aware of the pattern, (iii) they have no plausible innocuous explanation for the pattern, (iv) they have no compelling reason to believe the pattern is not a problem, and (v) they chose to not tell you~\footnote{We use the phrase ``they chose to not tell you'' rather than ``they chose not to tell you'' deliberately.  The logical meaning is of course exactly the same and the latter is better grammar, but it is to our ear too passive.  Faced with a strikingly suspicious return pattern in the world's stock markets that nobody can innocuously explain, the default is to tell you.  Not telling you is a choice.}.

\section{Strikingly Suspicious Overnight and Intraday Returns\label{sec:StrikinglySuspicious}}

\begin{figure*}[t]
\includegraphics[width=7in]{suspicious_index_returns_log_scale}
\caption{\label{fig:SuspiciousReturns}The value of one unit of local currency invested in each of twenty-one major stock market indices, getting only overnight (blue) or intraday (green) returns.  The horizontal axis of each plot extends from January 1, 1990 to \lastDataDate.  Each vertical axis has logarithmic scale, units of local currency, a value of 1 where the blue and green curves start at left, and explicitly marked values where each blue and green curve ends at right.  Thus, for example, if you had invested \$1 in the S\&P~500 index (top left plot) on the first date for which data are available and had gotten only intraday returns (from market open to market close), you would have \lastDataDateSpecific{\$0.95} on \lastDataDate.  If you had gotten only overnight returns (from market close to the next day's market open), you would have \lastDataDateSpecific{\$17.48}.  The code used to make this figure is available at Ref.~\cite{thisArticleWebpage}.  Data are publicly available from Yahoo!~Finance~\cite{yahooFinance}.  A version of this plot with linear vertical scale is provided in Ref.~\cite{knuteson2020strikingly}.}
\end{figure*}

\begin{figure*}[t]
\includegraphics[width=7in]{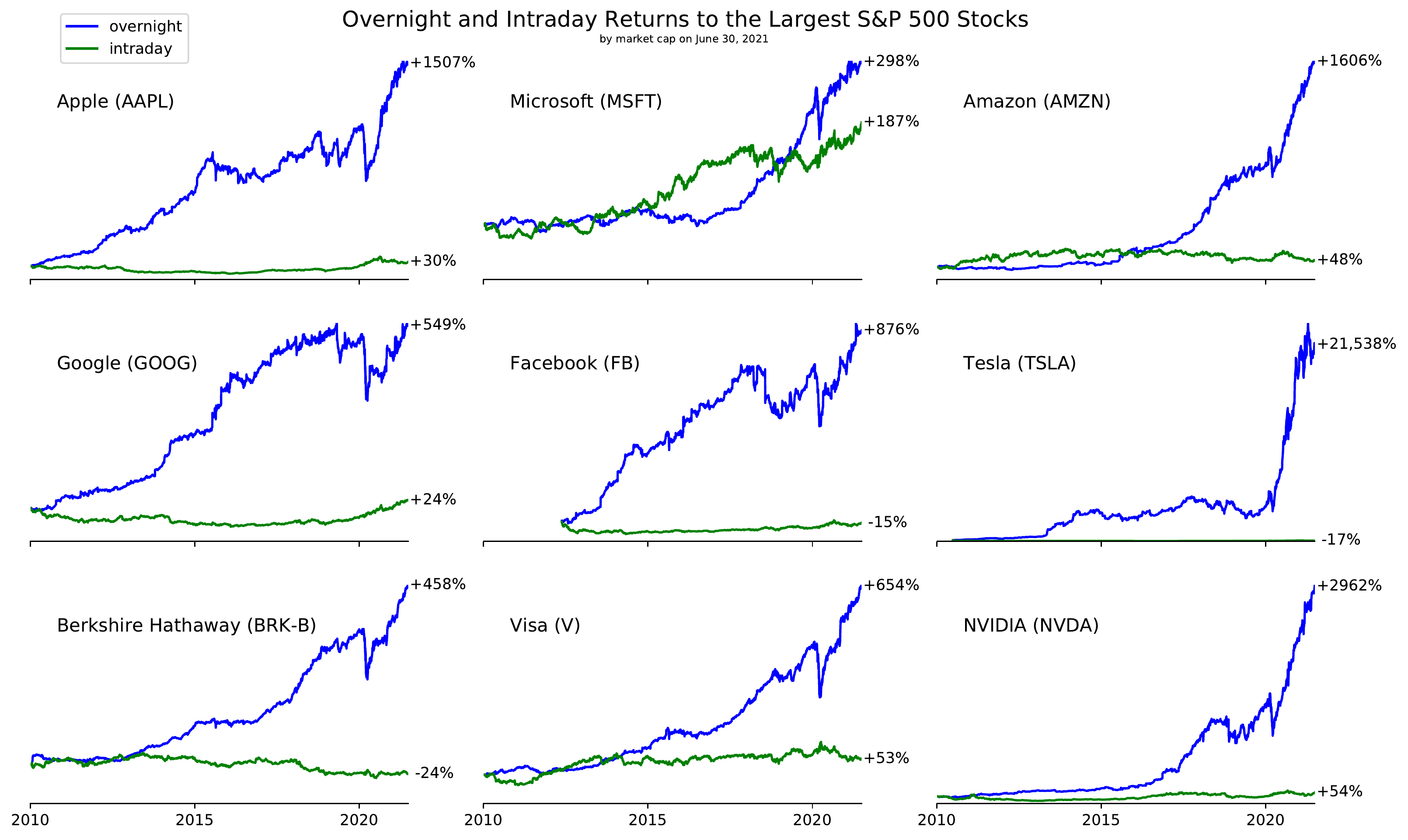}
\caption{\label{fig:SuspiciousReturnsLargestUSCompanies}Cumulative overnight (blue) and intraday (green) returns to the largest companies in the S\&P~500 index (by market capitalization on \lastDataDate) from the start of 2010 to \lastDataDate.  The (linear) vertical scale in each plot extends from a return of -100\% (bottom of plot) through 0 (where the blue and green curves start, at left) to the largest cumulative overnight return achieved (top of plot).  On each plot, the cumulative overnight and intraday returns on \lastDataDate\ are explicitly marked, at right.  Microsoft (top center) is in the ballpark of reasonable.  The rest are not.  The code used to make this figure is available at Ref.~\cite{thisArticleWebpage}.  Data are publicly available from Yahoo!~Finance~\cite{yahooFinance}.}
\end{figure*}


The world's major stock market indices display a mysterious, decades long, strikingly suspicious pattern of overnight and intraday returns.  \suspiciousReturnFigure\ shows the value of one unit of local currency invested in each of twenty-one major stock market indices around the world, getting only intraday (from market open to market close) or overnight (from market close to the next day's market open) returns.  For example, if you had invested \$1 in the NASDAQ Composite index at the start of 1990 and received only intraday returns, by \lastDataDateSpecific{the end of June 2021} you would have \lastDataDateSpecific{\$0.70} (top middle plot of \suspiciousReturnFigure), suffering a three-decades-long cumulative return of \lastDataDateSpecific{$-30\%$}.  \suspiciousReturnFigure\ contains the same information as a plot of cumulative overnight and intraday returns, shown in linear scale for these same indices in Ref.~\cite{knuteson2020strikingly}.  The extraordinary return patterns in \suspiciousReturnFigure\ are robust, undisputed, well documented in the academic literature~\citeAllOvernightIntradayLiterature, and easily reproducible.

It is important you appreciate just how simple \suspiciousReturnFigure\ is.  Open and close prices are among the most basic of financial data.  Making a plot of cumulative overnight and intraday returns is one of the very few things you can do with these data.  Having made such a plot (\suspiciousReturnFigure), you do not need to have been a particle physicist, MIT physics professor, and quant at a long-lived and incredibly consistently well performing hedge fund~\cite{wigglesworth2019deshaw} to appreciate that this plot should not look like it does.  The ability to distinguish lines that go up from lines that go down is more than sufficient.  If such an obvious problem can go unrecognized for so long, just think what other problems are being missed.

The problem we keep referring to is that the trading responsible for this suspicious overnight and intraday return pattern in the stock market over the past three decades is also likely responsible for the suspicious overall return in the stock market over the past three decades.  Consider Figure~\ref{fig:SuspiciousReturnsLargestUSCompanies}, which shows the presence of this remarkable overnight and intraday return pattern in eight of the nine largest companies (by market capitalization on \lastDataDate) in the S\&P~500 index over the past decade.  The presence of this extraordinary pattern in our largest companies is what you would expect to see if the trading that caused this extraordinary pattern also caused these companies to be our largest.  Alternatively, take a look at Tesla (middle right plot of Figure~\ref{fig:SuspiciousReturnsLargestUSCompanies}).  The total return to Tesla is suspicious, and the fact that more than all of it came overnight is suspicious.  These two suspicious things might have completely different causes, but probably not.

\begin{figure}[thb]
\includegraphics[width=3.5in]{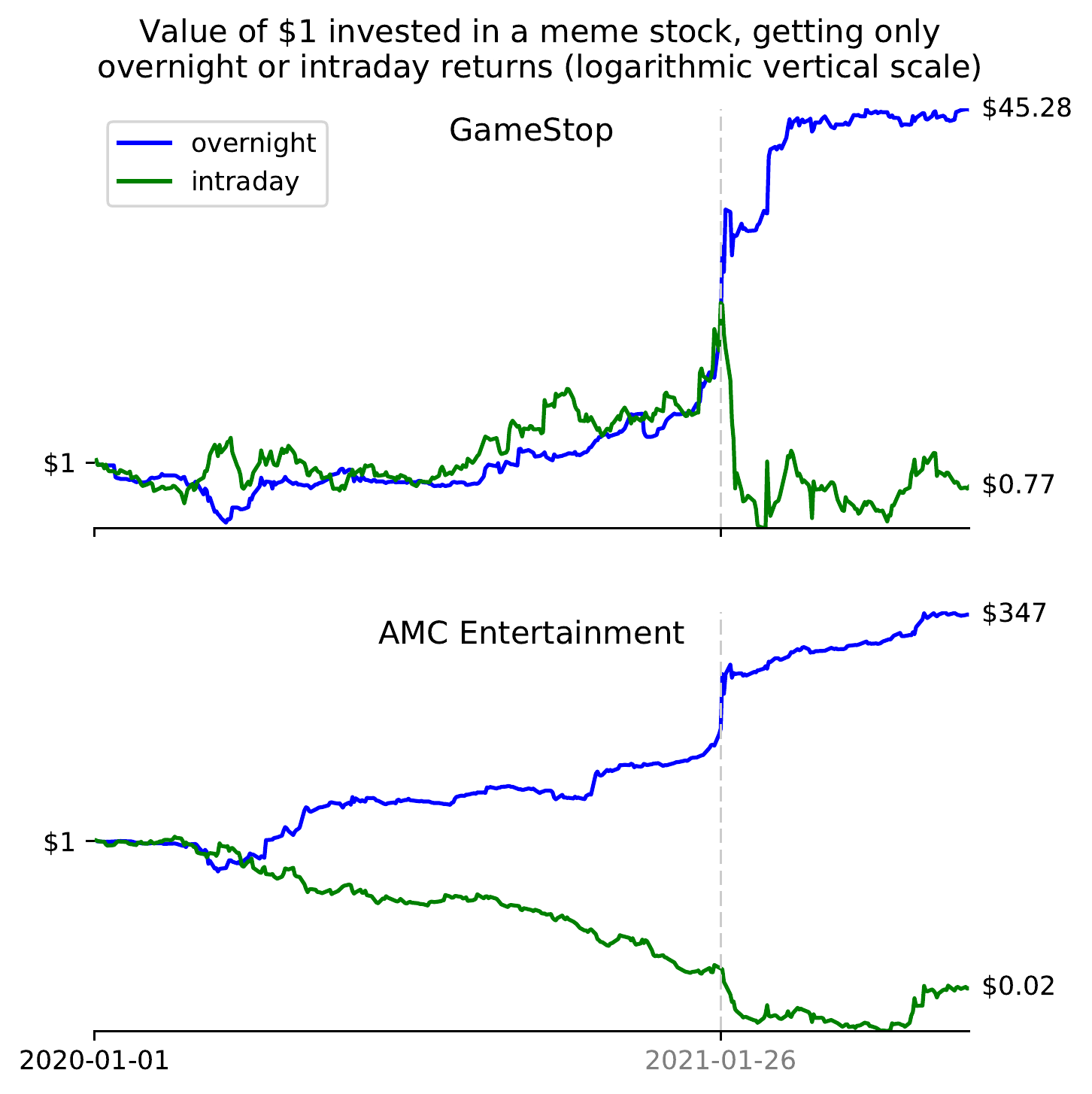}
\caption[Overnight/intraday returns to meme stocks]{\label{fig:SuspiciousReturnsMemeStocks}The value of \$1 of GameStop (top) and AMC (bottom) stock, invested at the start of 2020, getting only overnight (blue) or intraday (green) returns.  The vertical axis of each plot has logarithmic scale.  For example, \$1 invested in AMC at the start of 2020, getting only intraday returns, would leave you with \lastDataDateSpecific{\$0.02} on \lastDataDate, for a cumulative return of \lastDataDateSpecific{$-98\%$}.}
\end{figure}

Figure~\ref{fig:SuspiciousReturnsMemeStocks}, included purely for your amusement, shows overnight and intraday returns to two ``meme'' stocks~\cite{gensler2021meme}.

\begin{figure}[tb]
\includegraphics[width=3.5in]{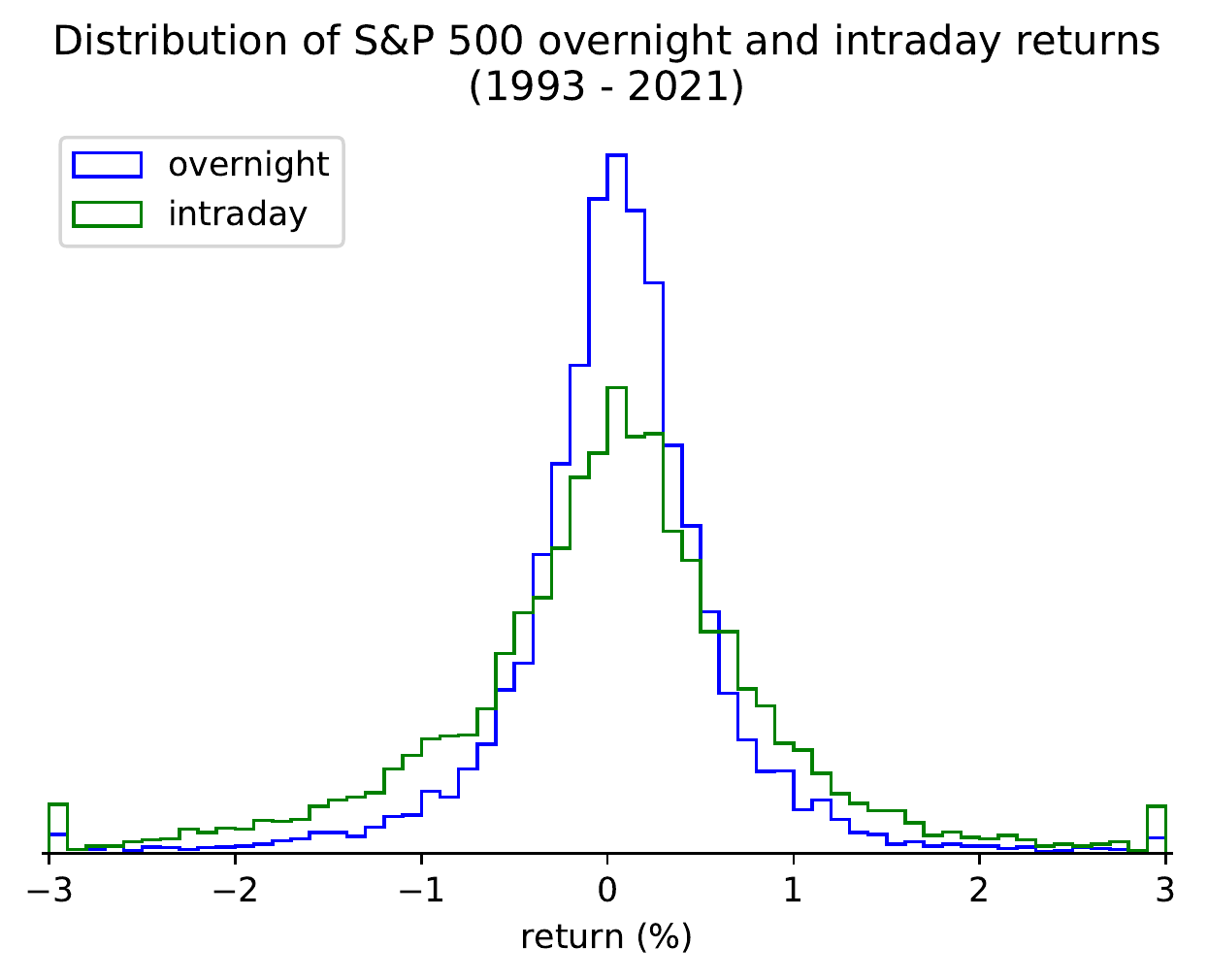}
\caption[Prices move more intraday]{\label{fig:PricesMoveMoreIntraday}Distribution of overnight and intraday returns to the S\&P~500 SPDR ETF from the first date for which data are available~\cite{yahooFinance} to \lastDataDate.  Each histogram has one entry for every trading day during this period.  Underflow and overflow are in the leftmost and rightmost bins, respectively.  The (green) distribution of intraday returns is clearly wider than the (blue) distribution of overnight returns.  For the S\&P~500 index and all other indices and stocks shown in this article over the time period we consider, prices move more intraday than they do overnight \ldots\ despite the many profoundly ignorant -- and, at the time of this writing, still uncorrected (e.g., Refs.~\cite{mccrum2018wrong,bloomberg2020margin,greifeld2020unbeatable}) -- public claims to the contrary.}
\end{figure}

Two salient features of \suspiciousReturnFigure\ have thwarted attempts to construct a plausible innocuous explanation~\cite{knuteson2020strikingly}:
\begin{enumerate}[itemsep=0.5em,parsep=-0.5em]
\item[(1)] the significantly negative intraday returns and
\item[(2)] the striking consistency of the plots.
\end{enumerate}
\vspace{-0.07in}
No contortion of ``returns are due to the bearing of risk''~\footnote{In many equity markets, roughly two-thirds of a typical 24-hour-day's price variance realizes intraday and one-third realizes overnight (Figure~\ref{fig:PricesMoveMoreIntraday}).  If returns were due to the bearing of risk, the intraday (green) returns in \suspiciousReturnFigure\ would be roughly twice the overnight (blue) returns, since two-thirds is twice one-third.} can plausibly explain (1), and the actions of millions of individual traders should not produce (2).  The nearly straight, diverging blue and green lines in \suspiciousReturnFigure\ look nothing like a random walk.

\setlength{\tabcolsep}{4.0em}
\begin{table}[b]
\caption[Plausible innocuous explanations]{\label{tbl:PlausibleInnocuousExplanations}A comprehensive list of all plausible innocuous explanations for the strikingly suspicious return patterns in \suspiciousReturnFigure\ in the public domain at the time of this writing.\\}
  \begin{tabular}{@{}ll@{}}
    \makecell{\bf Plausible Innocuous \\ \bf Explanation for \suspiciousReturnFigure} & \bf Reference \\ \hline
    ~\\~\\~\\
  \end{tabular}
\end{table}

\setlength{\tabcolsep}{5.3em}
\begin{table}[b]
\caption[Good reasons why everything is fine]{\label{tbl:PlausibleReasonsTheSuspiciousReturnsDoNotMatter}A comprehensive list of all convincing arguments in the academic literature for why the trading responsible for the strikingly suspicious return patterns in \suspiciousReturnFigure\ has not materially affected prices.\\}
  \begin{tabular}{@{}ll@{}}
    \makecell{\bf Good Reason Why \\ \bf \suspiciousReturnFigure\ is Fine} & \bf Reference \\ \hline
    ~\\~\\~\\
  \end{tabular}
\end{table}

We have previously explained~\cite{knuteson2020strikingly} how (1) and (2) can be used to straightforwardly assess the plausibility of attempted explanations for \suspiciousReturnFigure.  For example, you can discard the popular attempted explanation ``most price movement happens overnight'' simply by noting that this does not explain (1).  (For extra credit, you can disprove the premise by comparing the distribution of overnight returns to the (wider) distribution of intraday returns, as in Figure~\ref{fig:PricesMoveMoreIntraday}.  Prices move more intraday than they do overnight.)  You can discard what has become informally known as ``the New York Fed's explanation'' -- through no fault of the authors, most of whom are in Copenhagen -- by actually reading Ref.~\cite{boyarchenko2021overnight} and noting none of its 68 pages claim an explanation for (1), (2), or any other feature of \suspiciousReturnFigure~\footnote{Refs.~\cite{boyarchenko2021overnight,bondarenko2021market} observe and attempt to explain an interesting (and potentially related) puzzle in the S\&P~500 futures market.  Neither claims to explain even the top left plot of \suspiciousReturnFigure, let alone the twenty other plots in \suspiciousReturnFigure.}.  Applying similarly straightforward and elementary logic to all non-nefarious explanations for \suspiciousReturnFigure\ so far advanced~\cite{knuteson2020strikingly} leaves us with the list of plausible innocuous explanations shown in Table~\ref{tbl:PlausibleInnocuousExplanations}~\footnote{``To the best of our knowledge'' is of course implied everywhere.  We have followed the (English) literature on this topic closely and consider unlikely the possibility that any item deserving inclusion in Table~\ref{tbl:PlausibleInnocuousExplanations} or Table~\ref{tbl:PlausibleReasonsTheSuspiciousReturnsDoNotMatter} has escaped our notice.  The combined knowledge of those we consulted for this article obviously only partially covers the broad scope of Table~\ref{tbl:SuspiciousReturnsThatWereFine} and yet broader scope of Table~\ref{tbl:InstitutionsThatVoluntarilyAdmittedSignificantFailure}, so our having missed an item deserving inclusion in these tables is more likely.}.

Although understanding the cause of \suspiciousReturnFigure\ would obviously be helpful in determining whether \suspiciousReturnFigure\ is a problem, there could be good reasons to think that \suspiciousReturnFigure\ is not a problem, even if we do not definitively know the cause.  In this spirit, Table~\ref{tbl:PlausibleReasonsTheSuspiciousReturnsDoNotMatter} lists all good reasons provided in the academic literature for why the suspicious return patterns in \suspiciousReturnFigure\ are not a problem.

If you want a decent, unbiased estimate of the probability that \suspiciousReturnFigure\ is fine, an excellent way to go about it is to ignore all details of \suspiciousReturnFigure\ and simply divide the number of strikingly suspicious return patterns in financial markets that turned out to be fine by the number of strikingly suspicious return patterns in financial markets whose ultimate fineness (or not) has been determined~\cite{tetlock2015superforecasting}.  Table~\ref{tbl:SuspiciousReturnsThatWereFine} lists all historical examples of strikingly suspicious return patterns in financial markets that turned out to be fine.

\setlength{\tabcolsep}{1.5em}
\begin{table}[b]
\caption[Historical suspicious returns that turned out to be fine]{\label{tbl:SuspiciousReturnsThatWereFine}A list of all strikingly suspicious return patterns in financial markets throughout history that turned out to be just fine.\\}
  \begin{tabular}{@{}lll@{}}
    \makecell{\bf Suspicious Return \\ \bf that turned out \\ \bf to be Fine } & \makecell{\bf Totally OK \\ \bf Cause} & \bf Reference \\ \hline
    ~\\~\\~\\
  \end{tabular}
\end{table}

A similar approach can be used to estimate the probability that a financial regulator will investigate and publicize the cause of \suspiciousReturnFigure, given that \suspiciousReturnFigure\ is indeed a problem~\footnote{The phrase ``given that'' indicates a conditional probability -- the probability of something assuming (conditioned on) something else.  In this case, we ask:  If \suspiciousReturnFigure\ is indeed a problem, what is the probability a financial regulator will investigate and publicize its cause?  Because this probability is small (as argued in the main text), the fact that no financial regulator appears to have investigated and publicized the cause of \suspiciousReturnFigure\ is only weak evidence that everything is fine.}.  The United States Securities and Exchange Commission's stated mission is to protect investors, maintain fair, orderly, and efficient markets, and facilitate capital formation.  Many other financial watchdogs have similar stated missions.  If \suspiciousReturnFigure\ is indeed a problem, then investigating and publicizing the cause of \suspiciousReturnFigure\ is tantamount to the regulator voluntarily admitting it has stunningly failed in its core mission for decades.  A reasonable estimate of the probability that a financial regulator will investigate and publicize the cause of \suspiciousReturnFigure, given that \suspiciousReturnFigure\ is indeed a problem, can therefore be obtained simply by dividing the number of institutions throughout history that voluntarily admitted (without external forcing) that they failed in their core mission for decades by the number of institutions throughout history that failed in their core mission for decades.  Table~\ref{tbl:InstitutionsThatVoluntarilyAdmittedSignificantFailure} lists all historical examples of institutions that voluntarily admitted they failed in their core mission for decades.

\setlength{\tabcolsep}{2em}
\begin{table}[b]
\caption[Institutions that voluntarily admitted they failed]{\label{tbl:InstitutionsThatVoluntarilyAdmittedSignificantFailure}A list of all institutions throughout history that voluntarily admitted they failed in their core mission for decades.\\}
  \begin{tabular}{@{}lll@{}}
    \bf Institution & \makecell{\bf Stated \\ \bf Mission} & \makecell{\bf Date of Voluntary  \\ \bf  Admission of \\ \bf Mission Failure} \\ \hline
    ~\\~\\~\\
  \end{tabular}
\end{table}

Briefly summarizing the facts we have established so far: (i) the world's stock markets display a stunning, robust, undisputed, easily reproducible pattern of overnight and intraday returns (\suspiciousReturnFigure), (ii) the academic literature contains exactly zero plausible innocuous explanations for \suspiciousReturnFigure\ (Table~\ref{tbl:PlausibleInnocuousExplanations}),  (iii) the academic literature contains exactly zero good reasons for why \suspiciousReturnFigure\ is not a problem (Table~\ref{tbl:PlausibleReasonsTheSuspiciousReturnsDoNotMatter}), and (iv) there are very good reasons to think \suspiciousReturnFigure\ is indeed a problem (including Figures~\ref{fig:SuspiciousReturnsLargestUSCompanies} and~\ref{fig:SuspiciousReturnsMemeStocks}, Table~\ref{tbl:SuspiciousReturnsThatWereFine}, and very basic common sense).

\section{They Chose to Not Tell You\label{sec:Chose}}

Almost everyone can distinguish blue lines that go up from green lines that go down, and almost everyone understands that strikingly suspicious return patterns in financial markets should be viewed as a problem until definitively shown otherwise.  Unfortunately, the public has been unable to apply such basic common sense to this issue because they are unaware of this issue because nobody has alerted them to it.  To firmly document the fact that nobody has alerted the public to this problem and justify our claim ``they did not tell you,'' we explicitly analyze the few articles that have come closest to doing so in Section~\ref{sec:Chose:Articles} and explain how each fails.  We justify our use of the word ``chose'' and our claim ``they chose to not tell you'' in Section~\ref{sec:Chose:Emails}.

\subsection{Articles \label{sec:Chose:Articles}}

We consider two pools of articles:  scholarly articles (mostly written by active academics) and news articles (mostly written by professional journalists).

Among the scholarly articles written by others on this topic, we particularly recommend Refs.~\cite{cooper2008return,lachance2015night}, which we credit as the first to clearly note the existence of these extraordinary return patterns in the United States and around the world, respectively.  The fact that neither of these articles was published tells you little about the quality of these articles and much about the quality of the academic publication process.  We also credit Ref.~\cite{qiao2020overnight} with pointing out a particular rule unique to China's stock market that is plausibly responsible for China being the exception in \suspiciousReturnFigure, although we differ on why this rule is important~\cite{knuteson2019celebrating}.  None of these articles nor any other academic article we are aware of has emphasized \suspiciousReturnFigure\ as a potentially serious problem.

Among the few news articles on this topic, we consider a February 2, 2018 article in \TheNYT\ (``The Stock Market Works by Day, But It Loves the Night'')~\cite{sommer2018night} to be far and away the best due to its wide readership, its appropriate choice of people to quote, its prominent display of the top left plot in \suspiciousReturnFigure\ (with linear vertical scale), and its emphasis of the mystery behind the cause of this plot.  The top left plot in \suspiciousReturnFigure\ actually differs from the plot in \TheNYT\ because the latter throws dividends into the trash -- an odd choice, since throwing dividends into the trash is not the sort of thing people typically do with dividends -- thereby underestimating the cumulative overnight return to the S\&P~500 index by nearly a factor of two~\footnote{In the United States and other countries where dividends occur when the market is closed, how you treat those dividends affects the overnight (blue) curves and does not affect the intraday (green) curves at all.  Where our data~\cite{yahooFinance} allow (including the S\&P~500 index in \suspiciousReturnFigure\ and all stocks in Figures~\ref{fig:SuspiciousReturnsLargestUSCompanies} and~\ref{fig:SuspiciousReturnsMemeStocks}), we assume dividends are immediately reinvested (ignoring taxes, and thus slightly overstating the true overnight return).  Where our data do not allow (including the NASDAQ Composite index and many other indices in \suspiciousReturnFigure), we (like Ref.~\cite{sommer2018night}) effectively throw dividends into the trash and thus understate the true overnight return.}.  This article in \TheNYT\ also misleadingly implies the existence of plausible innocuous explanations when none in fact exist.  Such mistakes -- throwing dividends into the trash exactly like nobody ever does and implying the existence of plausible innocuous explanations where none exist -- are the level of intellectual rigor in the news article we consider to be the very best on this topic.  This article also nicely displays the typical ways such articles refrain from concluding that \suspiciousReturnFigure\ is a problem:  understating its scope (e.g., by focusing on the S\&P~500 index, the least obviously problematic plot in \suspiciousReturnFigure), making unjustifiable analysis choices that understate the magnitude of the problem (e.g., throwing dividends into the trash), and suggesting the plausibility of innocuous explanations that are clearly not plausible.

A \FT\ article~\cite{mccrum2018wrong} written a few days later (perfectly titled ``Someone is Wrong on the Internet, Day Versus Night Edition'') incorrectly claims the explanation for \suspiciousReturnFigure\ is that companies release good news, particularly earnings announcements, when the markets are closed.  (Ignoring returns around earnings announcements does not change \suspiciousReturnFigure\ at all~\cite{cooper2008return}.)  This and ``most news happens overnight'' are popular variants of ``most price movement happens overnight.''  See Figure~\ref{fig:PricesMoveMoreIntraday}.

A September 17, 2020 article in {\textit{Bloomberg News}} (``Unbeatable Overnight Gains Fuel Theories on Who's Driving Them'')~\cite{greifeld2020unbeatable} focuses solely on the S\&P~500 index, restricts the time period in its primary plot to only six months, and incorrectly implies that ``most price movement happens overnight'' and ``the New York Fed's explanation'' are plausible explanations.

There are a few other similar news articles with similar flaws, and that is basically it.  Zero news articles of wide readership have clearly alerted the public to the strikingly suspicious return patterns in \suspiciousReturnFigure\ as a potential problem.  A search of financial watchdogs' web pages and press releases similarly produces zero clear warnings about the strikingly suspicious overnight and intraday return patterns in the markets they are supposed to be watching.

To summarize:  They did not tell you.

\subsection{Emails \label{sec:Chose:Emails}}

We have repeatedly~\citeKnuteson\ pointed out that the accuracy of \suspiciousReturnFigure\ is not under dispute, that nobody has a plausible innocuous explanation for \suspiciousReturnFigure, and that nobody has a good reason to think \suspiciousReturnFigure\ is not a problem.  Perhaps showing will be more effective than telling.

We have chosen six email threads (out of hundreds on this topic over the past five years) that should sufficiently convey these points, and we have made these threads publicly available at Ref.~\cite{thisArticleWebpage}.  Three are with federal institutions in the United States -- the Securities and Exchange Commission, the Office of Financial Research (OFR), and the Federal Reserve Bank of New York -- chosen because of the role each of these institutions is expected to play in bringing potentially problematic financial issues to the public's attention.  The remaining three threads are with news institutions -- the \FT, \TheWSJ, and \TheWashPost\ -- chosen because they contain some of the most thoughtful comments and most insightful questions we have received from hundreds of journalists on this topic.

Pointing out a problem threatening the stability of the global financial system certainly justifies showing a few emails, however distasteful we find it, and there is clearly no problem with our doing so here in any case.  We have selected threads involving multiple recipients so each thread may be viewed as with the respective institution and have pruned tangential branches with individuals~\footnote{We are happy to restore any branch we have pruned at that person's explicit request.}.  All threads are fully on the record, most of the words are ours, and there is nothing improper, confidential, proprietary, or compromising among the words written by others.

Indeed, the deafening silence from the three federal institutions is of particular note.  If any knew the cause of \suspiciousReturnFigure\ and that cause was innocuous, they presumably would have simply stated the innocuous cause -- particularly when provoked with an audience, an opportunity we very deliberately provided in our thread with the OFR.  That they did not suggests they do not know the cause or they know the cause and that cause is not innocuous.  Neither of these scenarios is good.  When reviewing the threads with news institutions, bear in mind that journalists have little time, no technical expertise, and sources with huge conflicts of interest, so give them a break.

We show these six threads not because anything should be made out of any one of them, but simply because showing (rather than telling) may be the most effective way to convey the overall pattern.  Think of racial profiling: setting entirely to the side the question of whether any individual's choice can be argued as a justifiable judgment call, the overall pattern -- nobody alerting you to the potential problem suggested by the presence of strikingly suspicious return patterns in the world's stock markets that nobody seems able to innocuously explain -- is stomach-churningly disturbing.  Moreover, it is stomach-churningly disturbing even if \suspiciousReturnFigure\ somehow miraculously turns out to be fine.

If \suspiciousReturnFigure\ turns out to be the problem we expect, then the failure of our institutions~\footnote{Here and elsewhere, the general phrase ``our institutions'' refers to a large set of public and private institutions.} to bring this issue to your attention has even more in common with racial profiling.  Artificially increased stock prices increase the wealth of those who own stocks, and in the United States those people are disproportionately white.  If \suspiciousReturnFigure\ is the problem we describe in  Section~\ref{sec:StrikinglySuspicious}, then (employing a racial wealth inequality metric like the total wealth of whites divided by the total number of white people compared to the total wealth of Blacks divided by the total number of Black people) the failure of our institutions to recognize the problem so obvious in \suspiciousReturnFigure\ has likely contributed more to the current level of racial wealth inequality than any other act (or failure to act) during this time.

\section{Summary\label{sec:Summary}}

The first goal of this article was to alert you to a strikingly suspicious return pattern in the world's stock markets and show why the silence of others does not mean everything is fine.  None of the hundreds of people with whom we have corresponded on this topic have provided a plausible innocuous explanation for \suspiciousReturnFigure\ or a good reason for why \suspiciousReturnFigure\ is fine.  The silence of others hides an embarrassed ignorance and fear of looking foolish (among other things), not some deep, widespread understanding that everyone is just far too modest to share.  We set this first goal because a careful weighing of the evidence strongly suggests \suspiciousReturnFigure\ indicates a serious problem with clear implications for many important personal decisions, and we -- unlike those who have chosen to remain silent on this issue -- think you deserve to have this information as you make those decisions.

The second goal was to document an extraordinary failure of information transfer.  Regardless of whether \suspiciousReturnFigure\ turns out to be the problem it so clearly seems to be, you should have been alerted to this issue by people other than us.  You were not alerted, and that is a truly remarkable failure in the institutions and incentive systems we all rely on to be warned of potential problems.  Specifically, we have documented that (i) the world's stock markets display a stunning pattern of overnight and intraday returns, (ii) many of the people you reasonably rely on to bring such an issue to your attention (including financial regulators, journalists, and academic economists) are aware of the pattern, (iii) they have no plausible innocuous explanation for the pattern, (iv) they have no compelling reason to believe this suspicious return pattern is not a problem, and (v) they chose to not tell you.  We have made a small but representative subset of this documentation available with this article so you can assess these claims for yourself~\cite{thisArticleWebpage}.

\bibliography{chose}

\end{document}